\documentclass[a4paper]{PoS}%
\usepackage{amsfonts}
\usepackage{amsmath}
\usepackage{amssymb}
\usepackage{rsfs}
\usepackage{graphicx}

\providecommand{\U}[1]{\protect\rule{.1in}{.1in}}
\abstract{%
The dual superconductivity is believed to be a promising mechanism for quark
confinement. Indeed, what this picture is true has been confirmed in the
maximal Abelian (MA) gauge. However, it is not yet confirmed in any other
gauge and the MA gauge explicitly breaks color symmetry. To remedy this
defect, we propose to use our compact formulation of a non-linear change of
variables on a lattice. This formulation has succeeded to extract the magnetic
monopole with integer-valued magnetic charge in the gauge-invariant way. In
this talk, we present measurements of various correlation functions for the
operators constructed from the CFN variables in SU(2) Yang-Mills theory. Some
of our results reproduce previous results obtained in MA gauge, e.g.,
DeGrant-Toussaint monopole, infrared Abelian dominance and off-diagonal gluon
mass generation. These studies preserve color symmetry, in sharp contrast to
the conventional MA gauge. We argue the gauge fixing independence of these
results and the implications to quark confinement.
}

\FullConference{XXIVth International Symposium on Lattice Field Theory\\
July 23-28, 2006\\
Tucson, Arizona, USA}

\title{%
Gluon mass generation and infrared Abelian dominance in Yang-Mills theory%
{\protect{ \vspace{-5cm}%
\begin{flushright}%
\normalsize \parbox{5cm}{KEK Preprint 2006-39 \\ CHIBA-EP-162}% 
\end{flushright}%
\vspace{4cm}%
}}%
}

\author{\speaker{A.~Shibata}$^{a}$, S.~Ito$^{b}$, S.Kato$^{b}$ K.-I.~Kondo$^{d}$,
T.~Murakami$^{e}$, and T.~Shinohara$^{e}$\\\llap{$^a$}Computing Research Center, High Energy Accelerator Research
Organization (KEK), Tsukuba 305-0801, Japan\\\llap{$^b$}Nagano National College of Technology, 716 Tokuma, Nagano 381-8550, Japan\\\llap{$^c$}Takamatsu National College of Technology, Takamatu City 761-8058, Japan\\\llap{$^d$}Department of Physics, Faculty of Science, Chiba University, Chiba,
263-8522 Japan\\\llap{$^e$}Graduate School of Science and Technology, Chiba University, Chiba
263-8522, Japan }

\ShortTitle{Gluon mass generation and infrared Abelian dominance in Yang-Mills theory}
\begin{document}

\section{Introduction}

Quark confinement is still an unsolved and challenging problem in theoretical
particle physics. The dual superconductivity\cite{dualsuper} is believed to be
a promising mechanism for the vacuum of the non-Abelian gauge
theory\cite{YM54}. Indeed, the relevant data supporting the validity of this
picture have been accumulated by numerical simulations especially since 1990
and some of the theoretical predictions \cite{tHooft81,EI82} have been
confirmed by these investigations: infrared Abelian dominance \cite{SY90},
magnetic monopole dominance \cite{SNW94} and non-vanishing off-diagonal gluon
mass \cite{AS99} in the Maximal Abelian gauge \cite{KLSW87}, which are the
most characteristic features for the dual superconductivity. However, they are
not yet confirmed in any other gauge and the MA gauge explicitly breaks color
symmetry. To establish this picture in gauge invariant way, we need to answer
how to define and extract the \textquotedblleft Abelian part\textquotedblright%
\ $\mathbb{V}_{\mu}$ from the original non-Abelian gauge field $\mathcal{A}%
_{\mu}$ which is responsible for the area decay law of the Wilson loop
average. The conventional Abelian projection \cite{tHooft81} is too naive to
realize this requirement. At the same time, we must answer why the remaining
part $\mathbb{X}_{\mu}$ in the non-Abelian gauge field $\mathcal{A}_{\mu}$
decouple in the low-energy (or long-distance) regime.

We propose to use a non-linear change of variables (NLCV) which was called the
Cho-Faddeev-Niemi (CFN) decomposition\cite{DG79,Cho80,FN98,Shabanov99} to
remedy the defect of ordinary approaches.
\cite{KKMSS05:0,SCGT:latt05,KKMSSI06} We introduce a compact representation of
NLCV on a lattice. The naive decomposition presented at the last conference
was improved to extract the magnetic monopole with integer-valued magnetic
charge in the gauge-invariant way.\cite{IKKMSS06} Some of our results
reproduce previous results obtained in MA gauge, e.g., DeGrant-Toussaint
monopole, infrared Abelian dominance.

\section{Lattice CFN variables or NLCV on a lattice}

We propose a formulation of NLCV on a lattice It is a minimum
requirement that such a lattice formulation must reproduce the continuum
counterparts in the naive continuum limit. \ In the continuum formulation
\cite{Cho80,KMS05}, a color vector field $\vec{n}(x)=(n_{A}(x))$ $(A=1,2,3)$
is introduced as a three-dimensional unit vector field. In what follows, we
use the boldface to express the Lie-algebra $su(2)$-valued field, e.g.,
$\mathbf{n}(x):=n_{A}(x)T_{A}$, $T_{A}=\frac{1}{2}\sigma_{A}$ with Pauli
matrices $\sigma_{A}$ ($A=1,2,3$). Then the $su(2)$-valued gluon field (gauge
potential) $\mathbf{A}_{\mu}(x)$ is decomposed into two parts:
\begin{equation}
\mathbf{A}_{\mu}(x)=\mathbf{V}_{\mu}(x)+\mathbf{X}_{\mu}(x),\label{eq:decomp}%
\end{equation}
in such a way that the color vector field $\mathbf{n}(x)$ is covariant
constant in the background field $\mathbf{V}_{\mu}(x)$:
\begin{equation}
0=\mathcal{D}_{\mu}[\mathbf{V}]\mathbf{n}(x):=\partial_{\mu}\mathbf{n}%
(x)-ig[\mathbf{V}_{\mu}(x),\mathbf{n}(x)],\label{covariant-const}%
\end{equation}
and that the remaining field $\mathbf{X}_{\mu}(x)$ is perpendicular to
$\mathbf{n}(x)$:
\begin{equation}
\vec{n}(x)\cdot\vec{X}_{\mu}(x)\equiv2\mathrm{tr}(\mathbf{n}(x)\mathbf{X}%
_{\mu}(x))=0.\label{orthogX}%
\end{equation}
Here we have adopted the normalization $\mathrm{tr}(T_{A}T_{B})=\frac{1}%
{2}\delta_{AB}$. Both $\mathbf{n}(x)$ and $\mathbf{A}_{\mu}(x)$ are Hermitian
fields. This is also the case for $\mathbf{V}_{\mu}(x)$ and $\mathbf{X}_{\mu
}(x)$. By solving the defining equation (\ref{covariant-const}), the
$\mathbf{V}_{\mu}(x)$ and the $\mathbf{X}_{\mu}(x).$ are obtained in the
form:
\begin{align}
\mathbf{V}_{\mu}(x) &  =\mathbf{V}_{\mu}^{\parallel}(x)+\mathbf{V}_{\mu
}^{\perp}(x)=c_{\mu}(x)\mathbf{n}(x)-ig^{-1}[\partial_{\mu}\mathbf{n}%
(x),\mathbf{n}(x)],\label{Vdef}\\
\mathbf{X}_{\mu}(x) &  =-ig^{-1}\left[  \mathbf{n}(x),\mathcal{D}_{\mu
}[\mathbf{A}]\mathbf{n}(x)\right]  \label{Xdef}%
\end{align}
where the second term $\mathbf{V}_{\mu}^{\perp}(x):=-ig^{-1}[\partial_{\mu
}\mathbf{n}(x),\mathbf{n}(x)]=g^{-1}(\partial_{\mu}\vec{n}(x)\times\vec
{n}(x))_{A}T_{A}$ is perpendicular to $\mathbf{n}(x)$, i.e., $\vec{n}%
(x)\cdot\vec{V}_{\mu}^{\perp}(x)\equiv2\mathrm{tr}(\mathbf{n}(x)\mathbf{V}%
_{\mu}^{\perp}(x))=0$. Here it should be remarked that the parallel part
$\mathbf{V}_{\mu}^{\parallel}(x)=c_{\mu}(x)\mathbf{n}(x)$, $c_{\mu
}(x)=\mathrm{tr}(\mathbf{n}(x)\mathbf{A}_{\mu}(x))$ proportional to
$\mathbf{n}(x)$ can not be determined uniquely only from the defining equation
(\ref{covariant-const}), and the perpendicular condition of (\ref{orthogX})
determines $\mathbf{V}_{\mu}^{\parallel}(x)$ and remainder part $\mathbf{X}%
_{\mu}(x).$

On a lattice, on the other hand, we introduce the site variable $\mathbf{n}%
_{x}=n_{x}^{A}\sigma_{A}$ in addition to the original link variable $U_{x,\mu
}$ which is related to the gauge potential $\mathbb{A}_{x^{\prime},\mu}$:
\begin{equation}
U_{x,\mu}=\exp(-i\epsilon g\mathbb{A}_{x^{\prime},\mu},),\label{defU}%
\end{equation}
where  $\left(  x^{\prime},\mu\right)  =(x+\mu/2,\mu)$ stand for the midpoint
of the link\footnote{In general, the argument of the exponential in
(\ref{defU}) is the line integral of a gauge potential along a link from $x$
to $x+\mu$. We adopt this convention to obtain the naive continume limit of
$\mathcal{O}(\epsilon^{2}).$ Note also that we define a color vector field
$\mathbf{n}(x):=n_{A}(x)T_{A}$ in the continuum, while $\mathbf{n}_{x}:=$
$n_{x}^{A}\sigma_{A}$ on the lattice for convenience.}.

In what follows, we use the blackboard boldface to express the field
determined by the link variable. Note that $\mathbf{n}_{x}$ is Hermitian,
$\mathbf{n}_{x}^{\dagger}=\mathbf{n}_{x}$, and $U_{x,\mu}$ is unitary,
$U_{x,\mu}^{\dagger}=U_{x,\mu}^{-1}$. \ The link variable $U_{x,\mu}$ and the
site variable $\mathbf{n}_{x}$ transform under the gauge transformation II
\cite{KMS05} as
\begin{equation}
U_{x,\mu}\rightarrow\Omega_{x}U_{x,\mu}\Omega_{x+\mu}^{\dagger}=U_{x,\mu
}^{\prime},\quad\mathbf{n}_{x}\rightarrow\Omega_{x}\mathbf{n}_{x}\Omega
_{x}^{\dagger}=\mathbf{n}_{x}^{\prime}.\label{eq:Gtrans}%
\end{equation}

Suppose we have obtained a "link variable" $V_{x,\mu}$ and $X_{x,\mu}$ as a
group element of $G=SU(2)$ through
\begin{align}
V_{x,\mu} &  =\exp(-i\epsilon g\mathbb{V}_{x^{\prime},\mu}),\label{defV}\\
\ X_{x,\mu} &  =\exp(-i\epsilon g\mathbb{X}_{x,\mu}).\label{defX}%
\end{align}
$\mathbb{V}_{x^{\prime},\mu}$are related to the $su(2)$-valued background
field $\mathbb{V}_{x^{\prime},\mu}$ where $\mathbb{V}_{x^{\prime},\mu}$ is to
be identified with the continuum variable (\ref{Vdef}) and hence $V_{x,\mu}$
must be unitary $V_{x,\mu}^{\dagger}=V_{x,\mu}^{-1}$. A lattice version of
defining equation (\ref{covariant-const}) and (\ref{orthogX}) are given by
\begin{align}
&  D_{\mu}^{(\epsilon)}[\mathbf{V}]\mathbf{n}_{x}:=\epsilon^{-1}[V_{x,\mu
}\mathbf{n}_{x+\mu}-\mathbf{n}_{x}V_{x,\mu}]=0,\label{covariant-constL}\\
&  \mathrm{tr}(\mathbf{n}_{x}X_{x,\mu})=0.\label{orthogX-contL}%
\end{align}
The defining equation (\ref{covariant-constL}) needs a lattice covariant
derivative for an adjoint field. We adopt the midpoint evaluation of the
difference $\partial_{\mu}^{(\epsilon)}\mathbf{n}_{x}=\epsilon^{-1}%
[\mathbf{n}_{x+\mu}-\mathbf{n}_{x}]=\partial_{\mu}\mathbf{n}_{x^{\prime}}+$
$\mathcal{O}(\epsilon^{2})$, therefore the continuum covariant derivative for
the adjoint field up to $\mathcal{O}(\epsilon^{2})$ at midpoint:\footnote{The
term $\frac{i\epsilon}{2}\left\{  g\mathbb{V}_{x^{\prime},\mu},\partial_{\mu
}\mathbf{n}_{x^{\prime}}\right\}  $ is of the order $\mathcal{O}(\epsilon
^{2}),$ since $\mathbb{V}_{x^{\prime},\mu}$ in contimume limit is obtained as
eq(\ref{Vdef}) and $\partial_{\mu}\mathbf{n}_{x^{\prime}}\cdot\mathbf{n}%
_{x^{\prime}}=0+\mathcal{O}(\epsilon).$}
\[
\epsilon^{-1}[V_{x,\mu}\mathbf{n}_{x+\mu}-\mathbf{n}_{x}V_{x,\mu}%
]=\partial_{\mu}\mathbf{n}_{x^{\prime}}-ig[\mathbb{V}_{x^{\prime},\mu
},\mathbf{n}_{x^{\prime}}]-\frac{i\epsilon}{2}\left\{  g\mathbb{V}_{x^{\prime
},\mu},\partial_{\mu}\mathbf{n}_{x^{\prime}}\right\}  +\mathcal{O}%
(\epsilon^{2}).
\]
The derivative (\ref{covariant-constL}) obeys the correct transformation
property, i.e., the adjoint rotation on a lattice:
\[
D_{\mu}^{(\epsilon)}[\mathbf{V}]\mathbf{n}_{x}\rightarrow\Omega_{x}(D_{\mu
}^{(\epsilon)}[\mathbf{V}]\mathbf{n}_{x})\Omega_{x+\mu}^{\dagger},
\]
provided that the link variable $V_{x,\mu}$ transforms in the same way as the
original link variable $U_{x,\mu}$:
\begin{equation}
V_{x,\mu}\rightarrow\Omega_{x}V_{x,\mu}\Omega_{x+\mu}^{\dagger}=V_{x,\mu
}^{\prime}.\label{transV}%
\end{equation}
This is required from the transformation property of the continuum variable
$\mathbf{V}_{\mu}(x)$,\footnote{This indicates that $V_{x,\mu}$ in
(\ref{defV}) is considerd as the link variable whose argument of the
exponential is the line integral of a gauge potential along a link from $x$ to
$x+\mu.$} see \cite{KMS05}. Therefore, we obtain the desired condition between
$\mathbf{n}_{x}$ and $V_{x,\mu}$.%
\begin{equation}
\mathbf{n}_{x}V_{x,\mu}=V_{x,\mu}\mathbf{n}_{x+\mu}.\label{Lcc}%
\end{equation}
The defining equation (\ref{Lcc}) for the link variable $V_{x,\mu}$ is
form-invariant under the gauge transformation II, i.e., $\mathbf{n}%
_{x}^{\prime}V_{x,\mu}^{\prime}=V_{x,\mu}^{\prime}\mathbf{n}_{x+\mu}^{\prime}$.

A lattice version of the orthogonality equation (\ref{orthogX}) given by
equation(\ref{orthogX-contL}) or%
\begin{equation}
\mathrm{tr}(\mathbf{n}_{x}\exp\{-i\epsilon g\mathbb{X}_{x,\mu}\})=\mathrm{tr}%
(\mathbf{n}_{x}\{\mathbf{1}-i\epsilon g\mathbb{X}_{x,\mu}\})+\mathcal{O}%
(\epsilon^{2})=0+\mathcal{O}(\epsilon^{2}). \label{cond2}%
\end{equation}
This implies that the trace vanishes up to first order of $\epsilon$ apart
from the second order term. \ Note that $X_{x,\mu}$ is defined on the lattice
site and transforms in the same way as $\mathbf{n}_{x}$:
\begin{equation}
X_{x,\mu}\rightarrow\Omega_{x}X_{x,\mu}\Omega_{x}^{\dagger}=X_{x,\mu}^{\prime
}\,, \label{Xtransf}%
\end{equation}
so that orthogonality condition (\ref{orthogX-contL}) is gauge invariant.

Then, we proceed to solve the defining equation (\ref{Lcc}) for the link
variable $V_{x,\mu}$ and equation (\ref{orthogX-contL}) for the variable
$X_{x,\mu},$ and express it in terms of the site variable $\mathbf{n}_{x}$ and
the original link variable $U_{x,\mu}$, as is the case that the continuum
variable $\mathbf{V}_{\mu}(x)$ and $\mathbf{X}_{\mu}(x)$ are expressed in
terms of $\mathbf{n}(x)$ and $\mathbf{A}_{\mu}(x)$. Remembering the relation
$\mathbb{X}_{x,\mu}=\mathbb{A}_{x,\mu}-\mathbb{V}_{x,\mu}$, $X_{x,\mu}$ can be
defined using link variables $V_{x,\mu}$ and $U_{x,\mu}$ contacting to site
$x$, and linear combination of $V_{x-\mu,\mu}^{\dagger}U_{x-\mu,\mu}$ and
$U_{x,\mu}V_{x,\mu}^{\dagger},$ are candidate to satisfy the required
transformation property (\ref{Xtransf});
\begin{align}
X_{x,\mu} &  =\lambda V_{x-\mu,\mu}^{\dagger}U_{x-\mu,\mu}+\phi U_{x,\mu
}V_{x,\mu}^{\dagger}\label{defX2}\\
&  =\exp(-i\epsilon g\mathbb{X}_{x,\mu})\left[  \lambda+\phi+(\lambda
-\phi)\frac{g^{2}\epsilon^{2}}{2}[\mathbb{V}_{x},\mathbb{A}_{x,\mu
}]+\mathcal{O}(\epsilon^{3})\right]  ,\nonumber
\end{align}
where the relation for matrices $\exp(\epsilon A)\exp(\epsilon B)=\exp
(\epsilon A+\epsilon B+\epsilon^{2}[A,B]/2+\mathcal{O}(\epsilon^{3}))${}${}$
and its inverted version of $\exp(\epsilon C+\epsilon^{2}D)=\exp(\epsilon
C)\exp(\epsilon^{2}D)+\mathcal{O}(\epsilon^{3})~$are used. The parameter
$\lambda=\phi$ $\ $is selected so that $X_{x,\mu}$ is determined to coincide
with continuum expression up to $\mathcal{O}(\epsilon^{3})$.

As for $V_{x,\mu}$, on the other hand, the equation (\ref{Lcc}) is a matrix
equation and it is rather difficult to obtain the general solution. Therefore,
we adopt an ansatz (up to quadratic in $\mathbf{n}$):
\begin{equation}
V_{x,\mu}=U_{x,\mu}+\alpha\mathbf{n}_{x}U_{x,\mu}+\beta U_{x,\mu}%
\mathbf{n}_{x+\mu}+\gamma\mathbf{n}_{x}U_{x,\mu}\mathbf{n}_{x+\mu
},\label{Vanzats}%
\end{equation}
which enjoys the correct transformation property, the adjoint rotation
(\ref{transV}). It turns out that this ansatz satisfy the defining equation
(\ref{Lcc}), if and only if the numerical coefficients $\alpha,\beta$ and
$\gamma$ are chosen to be $\gamma=1$ and $\alpha=\beta.$ Then, substituting
the ansatz (\ref{Vanzats}) with a still undetermined parameter $\alpha$ into
equation (\ref{defX2}), we obtain $\alpha=0+\mathcal{O}(\epsilon^{2})$ (see
\cite{KKMSSI06}).

Thus we have determined $V_{x,\mu}$ and $X_{x,\mu}$ up to an overall
normalization
\begin{align*}
V_{x,\mu}  &  =V_{x,\mu}[U,\mathbf{n}]=U_{x,\mu}+\mathbf{n}_{x}U_{x,\mu
}\mathbf{n}_{x+\mu}\,,\\
X_{x,\mu}  &  =X_{x,\mu}[U,\mathbf{n}]=V_{x-\mu,\mu}^{\dagger}U_{x-\mu,\mu
}+U_{x,\mu}V_{x,\mu}^{\dagger}\,.
\end{align*}
The unitary link variable $\hat{V}_{x,\mu}[U,\mathbf{n}]$ and $\hat{X}_{x,\mu
}[U,\mathbf{n}]$ can be obtained after the normalization:
\begin{equation}
\hat{V}_{x,\mu}[U,\mathbf{n}]:=V_{x,\mu}/\sqrt{\frac{1}{2}\mathrm{tr}%
[V_{x,\mu}^{\dagger}V_{x,\mu}]},\qquad\hat{X}_{x,\mu}[U,\mathbf{n}]:=X_{x,\mu
}/\sqrt{\frac{1}{2}\mathrm{tr}[X_{x,\mu}^{\dagger}X_{x,\mu}]}.
\end{equation}

\section{Numerical simulations and generation of configuration of NLCV}

We generate configurations of link variables $\{U_{x,\mu}\}$ using standard
Wilson action. The numerical simulation are performed on $24^{4}$ lattice at
$\beta=2.3$, $2.4$, $2.5$ by thermalizing 15000 sweeps, and on $36^{4}$
lattice at $\beta=2.5$, $2.6$, $2.7$ by thermalizing 18000 sweeps. 200
configurations are obtained every 300 sweeps. \begin{figure}[ptb]
\begin{center}
\vspace{-1cm}%
\includegraphics[angle=-90, width=2.8in]{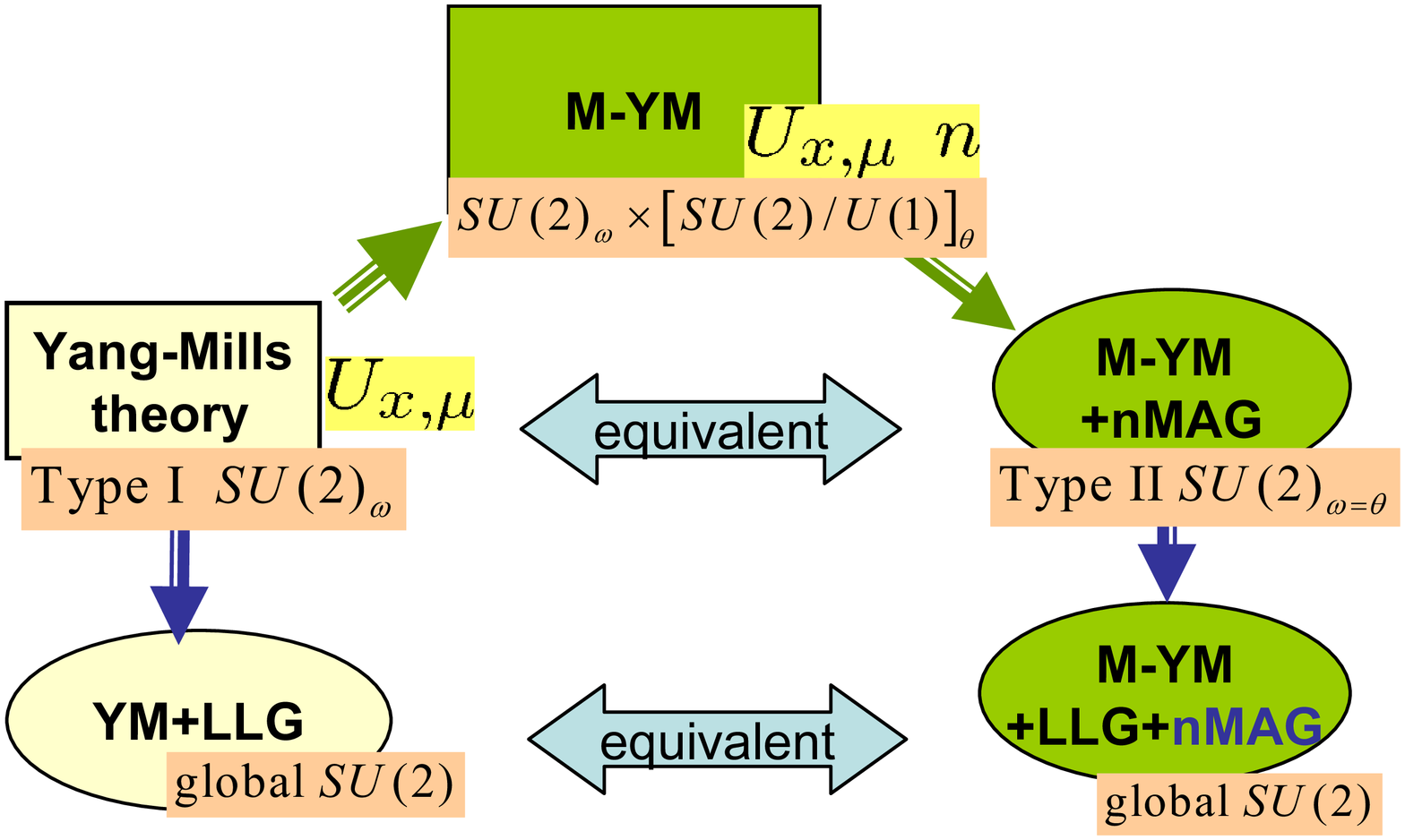}\includegraphics[angle=-90, width=2.8in]{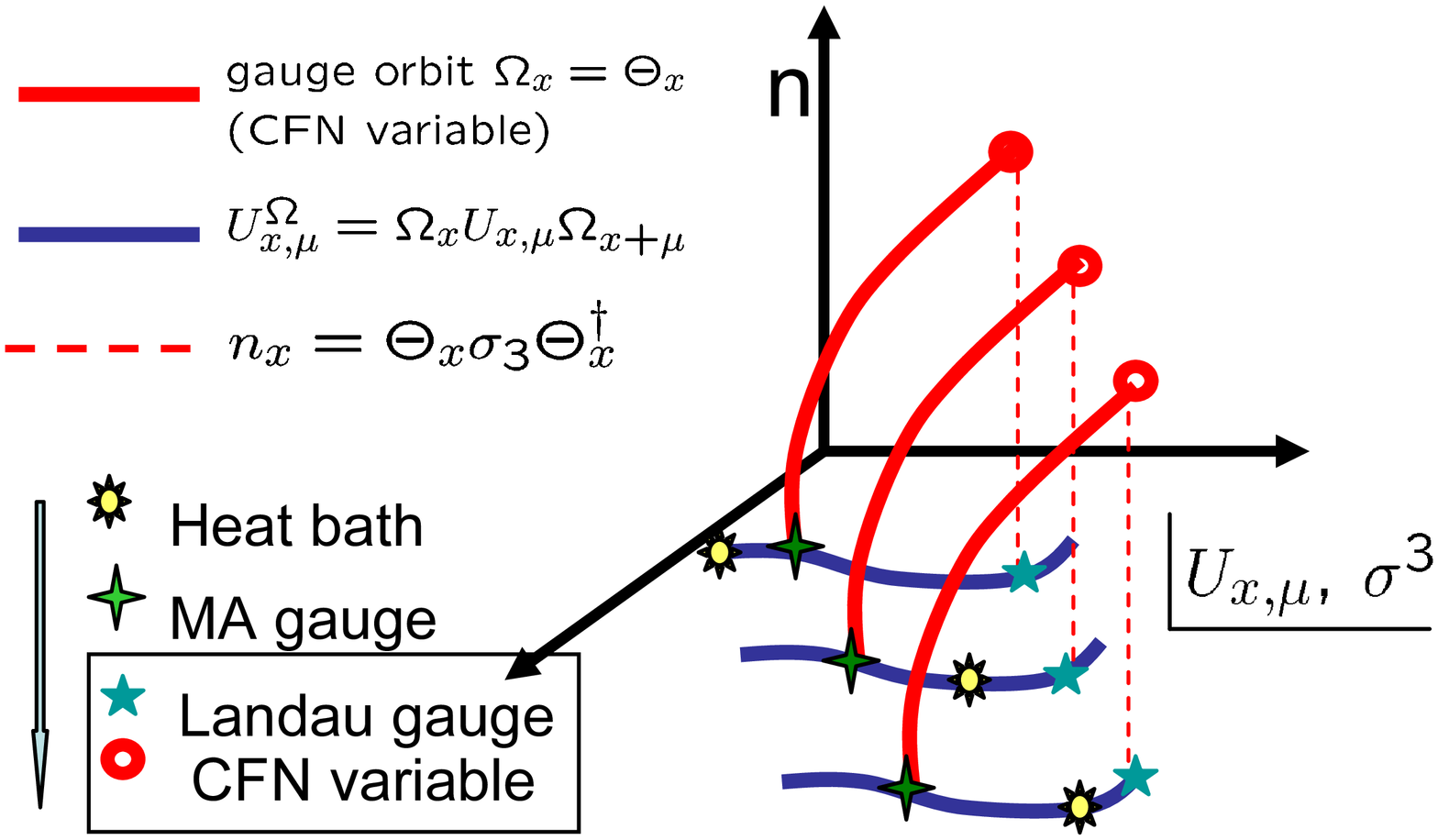}\vspace
{-1cm}
\end{center}
\caption{{}(left)The relationship between the master-YM theory and the
original YM\ theory. (right) NLCV via gauge transfroamtion. }%
\label{fig:M-YM&NLCV}%
\end{figure}

The NLCV on a lattice is obtained according to the method of the previous
paper \cite{KKMSSI06}. Figure \ref{fig:M-YM&NLCV} shows the extended gauge
symmetry in the master-YM for NLCV (left panel) and NLCV of SU(2) link
variables via gauge transformations (right panel). \ The configuration of
\ the link variable $U_{x,\mu}$ and the color vector field $\mathbf{n}_{x}$
has an extended gauge symmetry $SU(2)_{\omega}\times\left[  SU(2)/U(1)\right]
_{\theta}$. The equivalent theory to the original YM theory is obtained by the
gauge fixing which we call the new Maximal Abelian gauge (nMAG). We define a
functional written in terms of the gauge (link) variable $U_{x,\mu}$ and the
color (site) variable $\mathbf{n}_{x}$; $F_{nMAG}[U,\mathbf{n};\Omega
,\Theta]\equiv\sum_{x,\mu}\mathrm{tr}(\mathbf{1}-{}^{\Theta}\mathbf{n}_{x}%
{}^{\Omega}U_{x,\mu}{}^{\Theta}\mathbf{n}_{x+\mu}{}^{\Omega}U_{x,\mu}%
^{\dagger})$, where we have introduced the enlarged gauge transformation:
${}^{\Omega}{}U_{x,\mu}:=\Omega_{x}U_{x,\mu}\Omega_{x+\mu}^{\dagger}$ for the
link variable $U_{x,\mu}$ and ${}^{\Theta}\mathbf{n}_{x}:=\Theta_{x}%
\mathbf{n}_{x}^{(0)}\Theta_{x}^{\dagger}$ for an initial site variable
$\mathbf{n}_{x}^{(0)}$. The gauge group elements $\Omega_{x}$ and $\Theta_{x}$
are independent SU(2) matrices on a site $x$. After imposing the nMAG, the
theory still has the local gauge symmetry $SU(2)_{local}^{\omega=\theta}$,
since the \textquotedblleft diagonal\textquotedblright\ gauge transformation
$\mathbf{\omega}=\mathbf{\theta}$ does not change the value of the functional
$F_{nMAG}[U,\mathbf{n};\Omega,\Theta]$. Therefore, the configuration of
$\mathbf{n}_{x}$ can not be determined at this stage. In order to determine
$\mathbf{n}_{x}$, we need to impose another gauge fixing or a choice of the
gauge of link variable $U_{x,\mu}$ for fixing $SU(2)_{\omega}$. The desired
color vector field $\mathbf{n}_{x}$ is constructed from the interpolating
gauge transformation matrix $\Theta_{x}$ by choosing the initial value
$\mathbf{n}_{x}^{(0)}=\sigma_{3}$ and $\mathbf{n}_{x}:=\Theta_{x}\sigma
_{3}\Theta_{x}^{\dagger}=n_{x}^{A}\sigma^{A}$, $n_{x}^{A}=\mathrm{tr}%
[\sigma_{A}\Theta_{x}\sigma_{3}\Theta_{x}^{\dagger}]/2$ ($A=1,2,3$) where
$\left\{  \Theta_{x}\right\}  $ are given by gauge transformations that
satisfy $U_{x,\mu}=\Theta_{x}U_{x,\mu}^{MAG}\Theta_{x+\mu}^{\dagger}.$ For
example, we choose the conventional Lorentz-Landau gauge or Lattice Landau
gauge (LLG) for this purpose. The LLG can be imposed by minimizing the
function $F_{LLG}[U;\Omega]=\sum_{x,\mu}\mathrm{tr}(\mathbf{1}-{}^{\Omega
}U_{x,\mu})$ with respect to the gauge transformation $\Omega_{x}$ for the
given link configurations $\{U_{x,\mu}\}$.

\section{Infrared Abelian Dominance and Mass generation of the off-diagonal
gluon}

\begin{figure}[ptb]
\begin{center}
\vspace{-16mm}%
\includegraphics[width=2.8in]{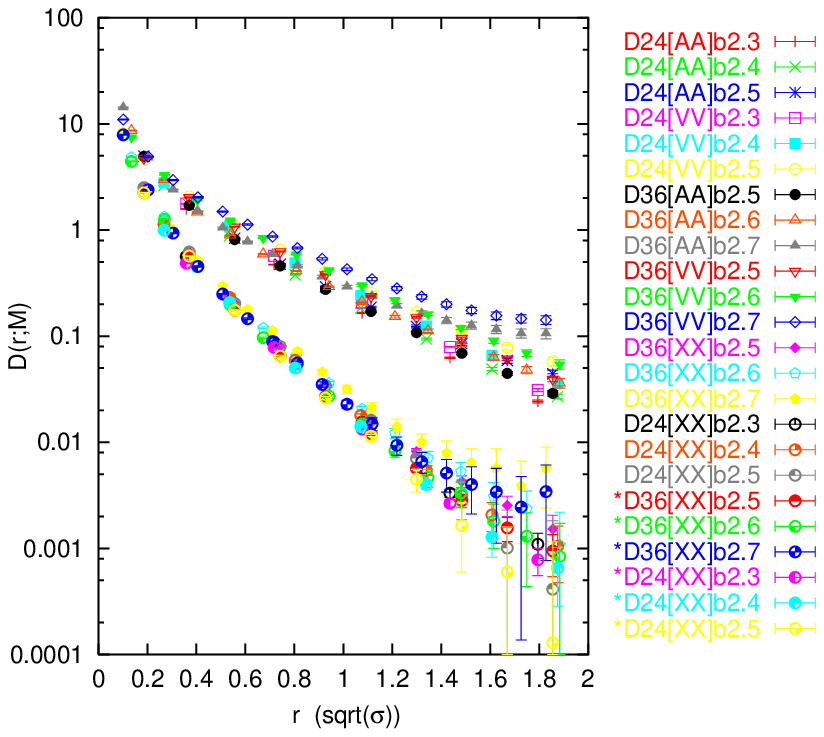}\includegraphics[width=2.8in]{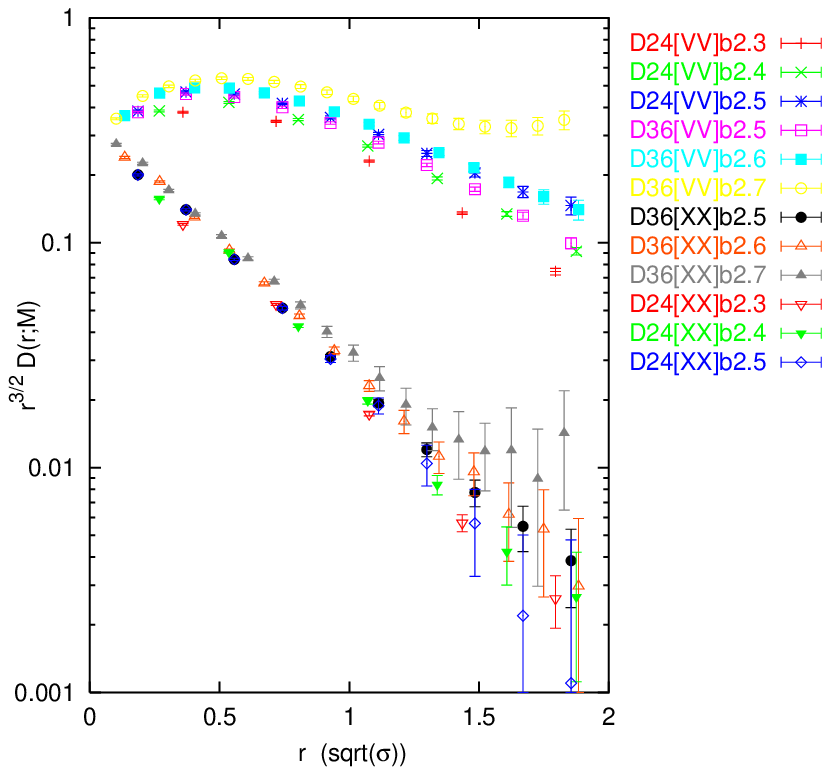}\vspace
{-8mm}
\end{center}
\caption{{}(left) correlation functions$\left\langle O(x)O(y)\right\rangle $
in the logalithmic scale, (right) rescaled correratin function $\ln\left(
r^{3/2}G_{\mu\mu}(r;M)\right)  $}%
\label{fig:prop}%
\end{figure}

Using new variables through NLCV, \ we are now ready to study characteristic
features of the YM\ theory for any choice of gauge fixing such as infrared
Abelian dominance, magnetic monopole dominance and the non-vanishing
off-diagonal gluon mass.\footnote{The magnetic monopole dominance has been
found using integer valued and gauge invariant magnetic monopole defined by
our NLCV. This fact has been reported in lattice2006 \cite{Lattice2006K} .}
Our proposed decomposition extract the \textquotedblleft Abelian
part\textquotedblright\ $V_{x,\mu}$ in any gauge fixing preserving the color
symmetry. The conventional MAG fixed theory is reproduced as a special case of
\ our formulation base on NLCV. To study the infrared Abelian dominance and
the non-vanishing off-diagonal gluon mass in LLG other than MAG, the
correlation function of the decomposed variable $V_{x,\mu}$ and $X_{x,\mu}$
has been measured. Left panel of figure \ref{fig:prop} shows propagators
$D_{AA}(x-y)=\left\langle \mathbb{A}_{x,\mu}\mathbb{A}_{y,\mu}\right\rangle $,
$D_{VV}(x-y)=\left\langle \mathbb{V}_{x,\mu}\mathbb{V}_{y,\mu}\right\rangle $
and $D_{XX}(x-y)=\left\langle \mathbb{X}_{x,\mu}\mathbb{X}_{y,\mu
}\right\rangle $. The gauge potentials are defined as link variables
$\mathbb{A}_{x^{\prime},\mu}=\frac{-i}{2g\varepsilon}\left[  A_{x,\mu
}-A_{x,\mu}^{\dagger}\right]  $, $\mathbb{V}_{x^{\prime},\mu}=\frac
{-i}{2g\varepsilon}\left[  V_{x,\mu}-V_{x,\mu}^{\dagger}\right]  .$ On the
other hand, we can define the $\mathbb{X}_{x,\mu}$ in two ways, one is
extracted from compact representation, $\mathbb{X}_{x,\mu}=\frac
{-i}{2g\varepsilon}\left[  X_{x,\mu}-X_{x,\mu}^{\dagger}\right]  ,$ and the
other is from definition of the decomposition (\ref{eq:decomp}),
$\mathbb{X}_{x^{\prime},\mu}=\mathbb{A}_{x^{\prime},\mu}-\mathbb{V}%
_{x^{\prime},\mu}.$ Plotting of two types of $D_{XX}(x-y)$ overlap for several
lattice spacings (several $\beta$s) , and the extracttion of the variable is
consistent (see left panel of figure \ref{fig:prop}). On the other hand,
$D_{AA}(x-y)$ and $D_{VV}(x-y)$ overlap, and $D_{XX}(x-y)$ is dumped more
quickly for infrared region than $D_{VV}(x-y)$. This implies that the infrared
Abelian dominance is found in the LLG. 

Next we study the mass of the decomposed fields from the correlation
functions. \ The inverse Fourier transformation of the massive gauge boson
propagator should behave for large $r=|x-y|$ as follows,
\[
G_{\mu\mu}(r;M)=\left\langle \mathbb{X}_{\mu}(x)\mathbb{X}_{\mu}%
(y)\right\rangle =\int\frac{d^{4}k}{(2\pi)^{4}}e^{ik(x-y)}\frac{1}{k^{2}%
+M^{2}}\left(  4+\frac{k^{2}}{M^{2}}\right)  \simeq\frac{3\sqrt{M}}%
{2(2\pi)^{3/2}}\frac{e^{-Mr}}{r^{3/2}}\,.
\]
So the scaled propagator $r^{3/2}G_{\mu\mu}(r;M)$ is proportional to $e^{-Mr}%
$, that is, the mass of gauge potential $M$, is obtained as the dumping factor
of $r^{3/2}G_{\mu\mu}(r;M)$. In other words, the gradient of the linear
fitting in the $r$ vs $\ln\left(  r^{3/2}G_{\mu\mu}(r;M)\right)  $ plot gives
the mass \thinspace\thinspace$M$. \ Right panel of figure \ref{fig:prop} shows
the plots of the scaled propagator of $\ \mathbb{X}_{x^{\prime},\mu}$ and
$\mathbb{V}_{x^{\prime},\mu}$. The distance $r$ is measured in the unit of
square root of the string tension $\sqrt{\sigma_{ST}}$ ($=440$ MeV), and
vertical axis is scaled in the logarithm to measure the dumping factor by the
linear fitting. To determine the physical scale, the relation between $\beta$
and lattice spacing $\epsilon$ is obtained from \cite{Kato}. The dumping of
propagator of $\mathbb{X}_{x,\mu}\,$ gives the mass $M_{X}\simeq1.18$ GeV, and
the \textquotedblleft Abelian part\textquotedblright\ $\mathbb{V}_{x,\mu}$
indicates $M_{V}\simeq0.48$ GeV. These are consistent with study in
MAG.\cite{AS99}

\section{Summary and discussion}

We have proposed a new formulation of \ the lattice Yang-Mills theory based on
the NLCV which was once called the CFN decomposition. This resolves all
drawbacks of the previous formulation of the decomposition on a lattice. This
compact formulation enables us to guarantee the magnetic charge quantization
in the gauge invariant way and to extract the \textquotedblleft Abelian
part\textquotedblright\ and the \textquotedblleft off-diagonal
part\textquotedblright\ preserving color symmetry \ in any choice of gauge of
the original YM theory. These features are sharp contrast to the conventional
MA gauge and these studies. We have measured the correlation function
(propagator in real space) in LLG. The Infrared Abelian dominance and the
gluon mass generation have been found. These results are consistent with study
in MA gauge.

\section*{Acknowledgments}

The numerical simulations have been done on a supercomputer (NEC SX-5) at
Research Center for Nuclear Physics (RCNP), Osaka University. This project is
also supported in part by the Large Scale Simulation Program No.06-17 (FY2006)
of High Energy Accelerator Research Organization (KEK). This work is
financially supported by Grant-in-Aid for Scientific Research (C) 18540251
from Japan Society for the Promotion of Science (JSPS), and in part by
Grant-in-Aid for Scientific Research on Priority Areas (B)13135203 from the
Ministry of Education, Culture, Sports, Science and Technology (MEXT).


\begin{thebibliography}{99}                                                                                               %


\bibitem {YM54}{\small C.N. Yang and R.L. Mills,
%Conservation of isotopic spin and isotopic gauge invariance,
Phys. Rev. 96, 191-195 (1954); R. Utiyama,
%Invariant theoretical interpretation of interaction,
Phys. Rev. 101, 1597-1607 (1956).}

\bibitem {dualsuper}{\small
%\bibitem{Nambu74}
Y. Nambu,
%Strings, monopoles, and gauge fields,
Phys. Rev. D 10, 4262
%--4268
(1974); G. 't Hooft, in: High Energy Physics, edited by A. Zichichi (Editorice
Compositori, Bologna, 1975); S. Mandelstam,
%Vortices and quark confinement in non-abelian gauge theories,
Phys. Report 23, 245
%--249
(1976); A.M. Polyakov,
%Compact gauge fields and the infrared catastrophe,
%Phys. Lett. B {\bf 59}, 82%--84
(1975).
%\bibitem{Polyakov77}
%A.M. Polyakov,
%Quark confinement and topology of gauge theories,
Nucl. Phys. B 120, 429
%--458
(1977).}

\bibitem {tHooft81}{\small G. 't Hooft,
%Topology of the gauge condition and new confinement phases in non-Abelian gauge theories,
Nucl.Phys. B190 [FS3], 455
%--478
(1981).}

\bibitem {EI82}{\small Z.F. Ezawa and A. Iwazaki,
%Abelian dominance and quark confinement in Yang--Mills theories,
Phys. Rev. D25, 2681
%--2689
(1982).}

\bibitem {SY90}{\small T. Suzuki and I. Yotsuyanagi,
%Possible evidence of abelian dominance in quark confinement,
Phys. Rev. D42, 4257
%--4260
(1990).}

\bibitem {SNW94}{\small J.D. Stack, S.D. Neiman and R. Wensley,
%String tension from monopoles in SU(2) lattice gauge theory,
[hep-lat/9404014], Phys. Rev. D50, 3399 (1994); H.~Shiba and T.~Suzuki,
Phys.Lett.B333, 461
%--
(1994).}

\bibitem {AS99}{\small K. Amemiya and H. Suganuma,
%Off diagonal gluon mass generation and infrared Abelian dominance in the maximally Abelian gauge in lattice QCD,
[hep-lat/9811035], Phys. Rev. D60, 114509 (1999); V.G. Bornyakov, M.N.
Chernodub, F.V. Gubarev, S.M. Morozov and M.I. Polikarpov,[hep-lat/0302002],
Phys. Lett. B559, 214-222 (2003).}

\bibitem {KLSW87}{\small A. Kronfeld, M. Laursen, G. Schierholz and U.-J.
Wiese,
%Monopole condensation and color confinement,
Phys.Lett. B 198, 516
%--520
(1987).}

\bibitem {GSZ01}{\small F.V. Gubarev, L. Stodolsky and V.I. Zakharov,
%On the significance of the vector potential squared,
[hep-ph/0010057], Phys. Rev. Lett. 86, 2220--2222 (2001).; F.V. Gubarev and
V.I. Zakharov,
%Emerging phenomenology of $\langle A^2_{min} \rangle$,
[hep-ph/0010096], Phys. Lett. B 501, 28--36 (2001).}

\bibitem {KMS05}{\small K.-I. Kondo, T. Murakami and T. Shinohara,
%Yang--Mills theory constracted from Cho--Faddeev--Niemi decomposition,
%Preprint CHIBA-EP-151,
[hep-th/0504107], Prog. Theor. Phys. 115, 201
%-216
(2006). K.-I. Kondo, T. Murakami and T. Shinohara,
%BRST quantization of the Yang--Mills theory in the Cho--Faddeev--Niemi decomposition,
[hep-th/0504198], Eur. Phys. J. C 42, 475
%--481
(2005).}

\bibitem {DG79}{\small Y.S. Duan and M.L. Ge, Sinica Sci., 11, 1072 (1979).}

\bibitem {Cho80}{\small Y.M. Cho,
%Restricted gauge theory,
Phys. Rev. D 21, 1080
%--
(1980).
%\\
%Y.M. Cho,
%Extended gauge theory and its mass spectrum,
Phys. Rev. D 23, 2415
%--
(1981).}

\bibitem {FN98}{\small L. Faddeev and A.J. Niemi,
%Partially dual variables in SU(2) Yang-Mills theory,
[hep-th/9807069], Phys. Rev. Lett. 82, 1624
%--
(1999).}

\bibitem {Shabanov99}{\small S.V. Shabanov,
%An effective action for monopoles and knot solitons in Yang-Mills theory,
[hep-th/9903223], Phys. Lett. B 458, 322
%--330
(1999). S.V. Shabanov,
%Yang-Mills theory as an Abelian theory without gauge fixing,
[hep-th/9907182], Phys. Lett. B 463, 263
%--272
(1999).}

\bibitem {KKMSS05:0}{\small S. Kato, K.-I. Kondo, T. Murakami, A. Shibata and
T. Shinohara,
%Numerical evidence for the existence of a novel magnetic condensation in Yang-Mills theory,
%Preprint, CHIBA-EP-150/KEK Preprint 2005-6,
hep-ph/0504054.}

\bibitem {SCGT:latt05}{\small A.Shibata S. Kato, S.Ito, K.-I. Kondo, T.
Murakami, A. Shibata and T. Shinohara, [hep-lat/0510027], PoS
LAT2005:332,2006}

\bibitem {KMS06}{\small K.-I. Kondo, T. Murakami and T. Shinohara,
%Yang--Mills theory constracted from Cho--Faddeev--Niemi decomposition,
%Preprint CHIBA-EP-151,
[hep-th/0504107], Prog. Theor. Phys. 115, 201
%-216
(2006). K.-I. Kondo, T. Murakami and T. Shinohara,
%BRST quantization of the Yang--Mills theory in the Cho--Faddeev--Niemi decomposition,
[hep-th/0504198], Eur. Phys. J. C 42, 475
%--481
(2005).}

\bibitem {KKMSSI06}{\small S. Kato, K.-I. Kondo, T. Murakami, A. Shibata, T.
Shinohara and S. Ito, [hep-lat/0509069], Phys. Lett. B 632, 326--332 (2006).}

\bibitem {IKKMSS06}{\small S. Ito, S. Kato, K.-I. Kondo, T. Murakami, A.
Shibata and T. Shinohara, [hep-lat/0604016].}

\bibitem {Lattice2006K}{\small S. Kato, S.Ito, K.-I. Kondo, T. Murakami, A.
Shibata and T. Shinohara, A talk in Lattice2006, to appear in the Proceedings
of Lattice2006, PoS Lattice(2006)068}

\bibitem {Kato}{\small S. Kato, S. Kitahara, N. Nakamura and T. Suzuki, Nucl.
Phys. B 520, 323-344 (1998).}
\end{thebibliography}
\end{document}